\documentclass[fleqn,10pt]{wlscirep}
\usepackage[utf8]{inputenc}
\usepackage[T1]{fontenc}
\title{Heterogeneous Update Processes Shape Information Cascades in Social Networks}

\author[1,*]{Flávio L. Pinheiro}
\author[2,3,4,+]{Vítor V. Vasconcelos}
\affil[1]{NOVA Information Management School (NOVA IMS) -- Universidade Nova de Lisboa, Portugal}
\affil[2]{Computational Science Lab, Informatics Institute, University of Amsterdam, Amsterdam, The Netherlands}
\affil[3]{POLDER, Institute for Advanced Study, University of Amsterdam, Amsterdam, The Netherlands}
\affil[4]{Center for Urban Mental Health, University of Amsterdam, Amsterdam, The Netherlands}

\affil[*]{fpinheiro@novaims.unl.pt}
\affil[+]{v.v.vasconcelos@uva.nl}
\begin{abstract}
    A common assumption in the literature on information diffusion is that populations are homogeneous regarding individuals' information acquisition and propagation process: Individuals update their informed and actively communicating state either through imitation (simple contagion) or peer influence (complex contagion). Here, we study the impact of the mixing and placement of individuals with different update processes on how information cascades in social networks. We consider Simple Spreaders, which take information from a random neighbor and communicate it, and Threshold-based Spreaders, which require a threshold number of active neighbors to change their state to active communication. Even though, in a population made exclusively of Simple Spreaders, information reaches all elements of any (connected) network, we show that, when Simple and Threshold-based Spreaders coexist and occupy random positions in a social network, the number of Simple Spreaders systematically amplifies the cascades only in degree heterogeneous networks (exponential and scale-free). In random and modular structures, this cascading effect originated by Simple Spreaders only exists above a critical mass of these individuals. In contrast, when Threshold-based Spreaders are assorted preferentially in the nodes with a higher degree, the cascading effect of Simple Spreaders vanishes, and the spread of information is drastically impaired. Overall, the study highlights the significance of the strategic placement of different roles in networked structures, with Simple Spreaders driving widespread cascades in heterogeneous networks and Threshold-based Spreaders playing a critical regulatory role in information spread with a tunable effect based on the threshold value. These effects have consequences to our understanding of social phenomena, such as the spread of innovations in heterogeneous social systems with the presence of eager (Simple Spreaders) versus averse (Threshold-based Spreaders) adopters, but also to information warfare on social media where Simple Spreaders can be seen as embedded agents (e.g., bots) used to amplify the virality of ill-intended content and, oppositely, Threshold-based Spreaders as an essential self-regulatory element of social systems operating as information filters.
\end{abstract}
\begin{document}

\flushbottom
\maketitle

\thispagestyle{empty}

\section*{\label{sec:intro} Introduction}
In the modern era of fast, global information flows, beyond the spread of innovations \cite{zhong2017linear,montanari2010spread} or opinions \cite{watts2007influentials,aral2012identifying,molaei2018information}, information wars and infodemics \cite{cinelli2020covid, prieto2021vaccination} present a scenario where individuals and algorithms with different adoption or learning profiles co-exist and spread information in complex networks. Intentionally, organizations can influence and sponsor individuals to propagate specific opinions, news, or facts, aiming at turning information viral and shaping public perceptions and views about varied topics \cite{kermani2017novel}. Indeed, organizations have used bot farms \cite{shao2017spread,shao2018spread,rossi2020detecting} or cyber armies \cite{bradshaw2017troops} embedded in the fabric of social media platforms, which are often indistinguishable from normal users \cite{shahid2022you}, to amplify the propagation of information or to keep platforms active. Network segregation has experimentally shown to favor the diffusion of false news \cite{stein2023network}. The effectiveness of such interventions underscores the importance of understanding the underlying dynamics of information diffusion on heterogeneous social networks and, particularly, the different roles individuals play.

A core assumption in past works is that the type of information being spread defines the type of adoption/learning process \cite{zheng2012social,cencetti2023distinguishing}. Consequently, it is tradition to consider homogeneous populations regarding individuals' learning profiles and spreading behaviors. For instance, individuals learn through contact (simple contagion \cite{zanette2002dynamics,dodds2004universal,smolyak2020mitigation,vespignani2012modelling,sood2008voter,sood2005voter}) or social pressure (complex contagion \cite{kempe2003maximizing,karsai2016local,centola2007cascade,centola2010spread,centola2018behavior,watts2007influentials,derechin2023cascades,borges2024social}). Naturally, all these models overlap in specific limits \cite{vasconcelos2019consensus} or interpretations and have their merits for addressing specific research questions, as they extend our understanding of the spread of different types of information---e.g., opinions, innovations, political affiliation, behaviors, habits---that we know to leave different distinguishable empirical traces \cite{berger2008identity,karsai2014complex,young2009innovation,sprague2017evidence}, and that may have different intervention points. 

However, such an assumption of homogeneity can be insufficient \cite{tump2020wise,mittal2024anti}. Indeed, in human-algorithm mixes, responses are expected to differ (i.e., between bots and humans). Even among humans, individuals with many friends/followers might not find it efficient to learn through simple imitation and, instead, rely on a general perception of the prevalence of opinions in their neighborhood to make decisions involving adopting new technology or behavior \cite{centola2018behavior,watts2007influentials}. In contrast, agents aiming to transmit specific information will share it as soon as they see it. From another viewpoint, some individuals might be more eager to adopt innovations in their neighborhood while others are more adverse to change and require pressure from multiple friends to adopt it \cite{robertson1967process,rogers2010diffusion,rogers2014diffusion}. 

Several studies have examined how individuals with varying susceptibility or influence can impede or facilitate diffusion processes. For instance, threshold-based agents have been explored primarily under homogeneous assumptions––focusing on either structural or behavioral effects in isolation\cite{kleinberg2007cascading, karampourniotis2015impact}. Further, most prior work considers static or random distributions of agent types, overlooking the strategic placement of different learner profiles within degree-heterogeneous networks. Recent advances suggest that network topology interacts strongly with node-level behaviors \cite{izquierdo2018mixing}, highlighting the need to investigate how high-degree nodes adopting stricter thresholds alter global diffusion patterns \cite{guilbeault2021topological}. And in the context of influence maximization literature much effort has been put into identifying best seeding strategies that maximize the unfolding cascades \cite{kim2014ct,tu2022viral,kempe2003maximizing,leskovec2007dynamics,aral2011commentary,aral2018social}

Given these examples, a more reasonable assumption is that populations are heterogeneous and contain individuals with different learning/adoption preferences/profiles  \cite{karampourniotis2015impact,mittal2024anti}. Such a co-existence of learning preferences creates rich dynamics in the irreversible adoption of information, including continuous and discontinuous phase transitions as the contact probability increases \cite{min2018competing}. 

Here, we consider the case of a heterogeneous population of \textit{Simple Spreaders} (individuals that activate through contact, \textit{i.e.}, simple contagion) and \textit{Threshold-based Spreaders} (whose adoption of information requires reinforcement from multiple peers, which we implement through a threshold function of the percentage of friends sharing the information, i.e., complex contagion). In that context, Simple Spreaders can be seen as individuals in the social fabric that promote the views of an external organization \cite{prier2020commanding,shao2017spread,ferrara2020misinformation}. We extend the literature by focusing on how the strategic placement of adoption profiles in different networks can regulate the spread of information. Understanding the strategic placement of agents can provide valuable insights into the effectiveness of information diffusion strategies, particularly in the context of information warfare.

Starting from a few seeds (i.e., the initial spreaders) placed randomly in the population, we explore how the size of an information cascade depends on the balance of Simple and Threshold-based Spreaders for different arrangements of these roles on social networks. We show that while scale-free networks facilitate larger cascade sizes for randomly placed roles in the population, they create obstacles when Threshold-based Spreaders preferentially occupy the network's most well-connected elements, meaning that cascades require many Simple Spreaders to unfold, an effect that is amplified on disassortative networks. Our results remain qualitatively consistent for increasing number of seeding nodes, suggesting a robust dynamical outcome and the role of Threshold-based spreaders as information filters.

\section*{\label{sec:model} Model \& Methods}
\subsection*{Social Structure}
Let us consider a population of $Z$ individuals whose social interactions are represented through a complex network. Individuals occupy the network nodes, while edges connect pairs of individuals and indicate the presence of a mutual social relationship of influence. As such, information spreads through social ties, i.e., the edges of the social network. The number of relationships an individual participates in defines their degree $k_i$, and $D(k)$ is the degree distribution that describes the relative frequency of individuals with degree $k$. The average degree of the population is $\langle k \rangle = \sum_k k D(k)$, which we fix across network types.

We consider three types of complex networks -- Random (ER), Exponential (Exp), and Scale-Free (SFBA) -- as the baseline structures. Random networks are generated using the Erdős–Rényi model \cite{erdHos1960evolution}, Scale-Free networks using the Barabási-Albert algorithm \cite{albert2002statistical} of growth and preferential attachment, and Exponential networks are generated following the growth algorithm of Barabási-Albert but with random attachment. These three networks provide three structures with low (ER), middle (Exp), and high (SFBA) levels of degree heterogeneity as measured by the variance of the degree distribution \cite{santos2012role}.

To generate degree Assortative and Disassortative variants of Scale-Free networks (SFBA) \cite{gleeson2008cascades}, we implement the Xulvi-Brunet \cite{xulvi2004reshuffling} algorithm with the constrain that only rewires that do not disconnect the network are accepted.

Finally, we generate a set of scale-free networks with different exponents through an algorithm that combines network growth with biased preferential attachment \cite{fortunato2006scale,goh2001universal,pinheiro2017intermediate} that proceeds as follows: starting from a clique of three nodes, $Z-3$ other nodes are added sequentially; each of the newly added nodes attaches to two pre-existing (resulting in a network with average degree of four) ones sampled proportionally to $t^{-\alpha}$ ($0.0 \leq \alpha \leq 1.0$) where $t$ corresponds to the ranked age of the nodes and $\alpha$ is a control parameter. The generated networks exhibit a power law degree distribution of the form $D(k) \approx k^{-\gamma}$, which in the limiting case of $Z \rightarrow \infty$ is defined by the control parameter $\alpha$ as $\gamma = (1 + \alpha)/\alpha$ \cite{fortunato2006scale,goh2001universal}. 

More importantly, and besides the relationship with $\gamma$, a more interesting interpretation of $\alpha$ is that it regulates the level of degree heterogeneity of the resulting networks as it exhibits both a one-to-one relationship with $\gamma$ and var($k$), so that a low (large) $\alpha$ is associated with a low (large) level of degree heterogeneity. We generate eleven sets of such networks within the range of $\alpha = 1/3$ to $\alpha = 1.0$ that exhibit exponents ranging between $\gamma \approx 3.30$ and $\gamma \approx 2.25$, respectively. The variance of the degree distribution of the highest degree heterogeneous network ($\alpha = 1.0$ and $\gamma = 2.25$) is $15$ times larger than that of the least heterogeneous one ($\alpha = 1.0$ and $\gamma = 2.25$).

To account for the stochasticity of the network generation models, we independently generated $100$ instances of each network type.

\subsection*{Cascade Dynamics}
Individuals can be in one of two states: active or inactive. Active means that they have already adopted the information being spread and can also spread it. Inactive means that individuals have the potential to adopt the information and are not able to spread it or influence their peers. We consider two types of individuals that characterize their update rule, or Learning Profile, when inactive: Simple Spreaders and Threshold-based Spreaders. (Inactive) Simple Spreaders copy the state of a random neighbor. (Inactive) Threshold-based Spreaders become active if the fraction of active neighbors is above $\Gamma$. 

At the start of each simulation, we consider every individual except for a small number of seeds inactive. Hence, the seeds correspond to a subset of active individuals that are responsible for the start of the spreading of (novel) information in the population. We consider seeds placed randomly in the population.

In contrast with past works \cite{jalili2017information,kleinberg2007cascading}, we assume individuals can update their state differently. In that sense, a fraction $\theta$ of individuals adopt a new state through pure imitation of random neighbors (\textit{i.e.}, simple contagion), Simple Spreaders. While the remaining fraction of individuals, $1-\theta$, requires reinforcement from multiple friends (\textit{i.e.}, complex contagion) to change their state, Threshold-based Spreaders.

\begin{figure*}[!t]
    \centering
    \includegraphics[width=\textwidth]{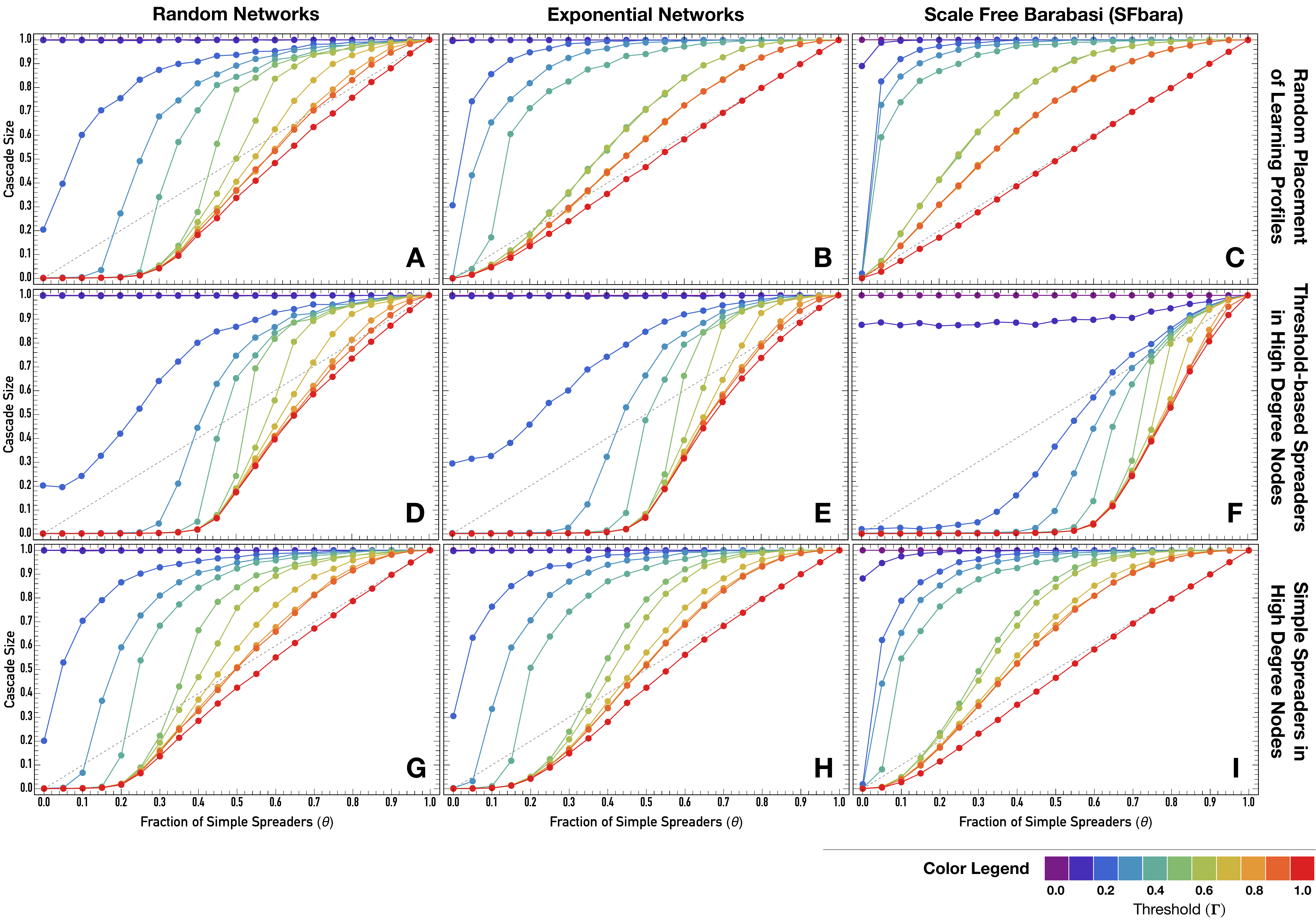}
    \caption{Cascade Sizes with one seeding individual when Learning Profiles are placed at random (Panels A to C) and when Threshold-based (Panels D to F) or Simple (Panels G to I) Spreaders are placed preferentially in high degree nodes. The diagonal dashed line represents the identity in which an increase in the fraction of Simple Spreaders ($\theta$) would lead to an equal increase in the final fraction of activated individuals (cascade size). Colors indicate different levels of the threshold used for Threshold-based Spreaders ($\Gamma$): 0.0 (Purple), 0.50 (green), and 1.00 (red). Other parameters are $Z = 10^3$ and $\langle k \rangle = 4$. For each condition, the reported results represent the average over $1000$ simulations for each of the $100$ network instances, totaling $10^5$ independent simulations.}
    \label{fig1}
\end{figure*}

We model the spreading dynamics as an asynchronous process in which, at each iteration, a random individual $i$ is selected to potentially update its state from inactive to active. If $i$ is already active, we skip to the next iteration, irrespective of the type. If $i$ is inactive and a Simple Spreader, then $i$ will sample a random neighbor $j$ and copy its state. This means that if $j$ is active, $i$ also becomes active; otherwise, it remains inactive. Finally, if $i$ is inactive and a Threshold-based Spreader, then $i$ will evaluate the fraction of active neighbors ($\gamma_i$), and, if $\gamma_i > \Gamma$, $i$ will become active. At each time step, there are $n$ active individuals and $Z-n$ inactive individuals in the population.

In each simulation, we repeat the update steps until the simulation either reaches i) a maximum of $Z\times10^6$ iterations, ii) the population reaches a monomorphic configuration where all individuals are active, or iii) the number of active individuals remains the same for $Z \times 10^2$ iterations. The final fraction of active individuals ($x = n/Z$) corresponds to the cascade size, which we average over $10^5$ independent simulations.

\section*{\label{sec:results} Results}
\subsection*{Analytical Insights}

To gain some intuition on this process, we consider the theory of percolation in infinite graphs. Our cascades can be analyzed in two distinct stages: the percolation of Simple Spreaders (SS) and the subsequent activation of Threshold-based Spreaders (TBS). This sequential analysis is rational because TBS will not activate spontaneously, and the formation of a giant connected component of SS serves as the backbone for triggering TBS activation through a bootstrap process. The general condition for percolation of SS nodes is that their average excess degree exceeds 1.
        
In Erdős-Rényi (ER) networks, this requires the fraction of SS-nodes, $\theta$, to be greater than $1/\langle k\rangle$, where $\langle k\rangle$ is the average degree of the network. Additionally, using a coarse mean-field approximation that assumes a homogenous degree, $\theta$ must exceed the threshold $\Gamma$ of TBS to ensure their activation, creating an additional constraint in $\theta$. 
        
Due to the high degree of heterogeneity in scale-free networks, we rely on heterogeneous mean-field approximations to describe the activation dynamics. The Mollow-Reed percolation condition still holds in its general form $\frac{\langle k(k-1)\theta\rangle}{\langle k \theta\rangle}>1$. When individuals are equally distributed across degrees, the fraction of SS-nodes per degree is $\theta=\theta$ for all degrees. Thus, it becomes clear that the percolation of SS-nodes depends on the degree-distribution exponent $\gamma$: for $\gamma \leq 3$, the numerator diverges, and the presence of hubs ensures that SS-nodes percolate even at very small $\theta$, while for $\gamma >3$, the condition requires a positive $\theta$. 
        
The condition for the percolation of TBSs following that of SSs requires numerical analysis. Degree-dependent assignment of a proportion $\theta_k$ of SS in a degree-$k$ class significantly influences the minimal fraction of SSs required for both SS- and TBS-node activation. A clear example is using a power-law relation for distributing SSs along degree, $\theta_k ~ k^{-\eta}$, where higher $\eta$ places more TBSs in higher degree nodes. Then, high enough $\eta$ can regularize the sum even for $\gamma\le 3$, preventing percolation of SSs for small fractions of SSs as long as $\eta>\gamma-3$. In what follows, we explore the cascade size systematically using simulation.

To gain some intuition on this process, we consider the theory of percolation in infinite graphs. Our cascades can be analyzed in two distinct stages: the percolation of Simple Spreaders (SS) and the subsequent activation of Threshold-based Spreaders (TBS). This sequential analysis is rational because TBS will not activate spontaneously, and the formation of a giant connected component of SS serves as the backbone for triggering TBS activation through a bootstrap process. The general condition for percolation of SS nodes is that their average excess degree exceeds 1.  \cite{molloy1995critical} 

In Erdős-Rényi (ER) networks, the first percolation of SS requires the fraction of SS-nodes, $\theta$, to be greater than $1/\langle k\rangle$, where $\langle k\rangle$ is the average degree of the network. Additionally, using a coarse mean-field approximation that assumes a homogenous degree, $\theta$ must exceed the threshold $\Gamma$ of TBS to ensure the activation of TBS, creating an additional constraint in $\theta$. 

As for scale-free networks, we rely on heterogeneous mean-field approximations to describe the activation dynamics due to their high degree of heterogeneity. The Mollow-Reed percolation condition still holds in its general form $\langle k(k-1)\theta\rangle / \langle k \theta\rangle>1$, where each class $k$ is characterized by a fraction of SS nodes $\alpha_k$. When individuals are equally distributed across degrees, the fraction of SS nodes per degree is $\theta_k = \theta$ for all degrees. Thus, it becomes clear that the percolation of SS nodes depends on the degree-distribution exponent $\gamma$: for $\gamma \leq 3$, the numerator diverges, and the presence of hubs ensures that SS nodes percolate even at very small $\theta$, while for $\gamma >3$, the condition requires a positive $\theta$. The condition for the percolation of TBSs following that of SSs requires numerical analysis. Degree-dependent assignments of a proportion $\theta_k$ of SS in a degree-$k$ class significantly influence the minimal fraction of SSs required for both SS- and TBS-node activation. A clear example is using a power-law relation for distributing SSs along degree, $\theta_k  \sim k^{-\eta}$, where higher $\eta$ places more TBSs in higher degree nodes (in the simulations that follow, we set $\eta=1$). Then, high enough $\eta$ can regularize the sum even for $\gamma\le 3$, preventing percolation of SSs for small fractions of SSs as long as $\eta>\gamma-3$. In what follows, we explore the cascade size systematically using simulations.

\begin{figure*}[!t]
    \centering
    \includegraphics[width=\textwidth]{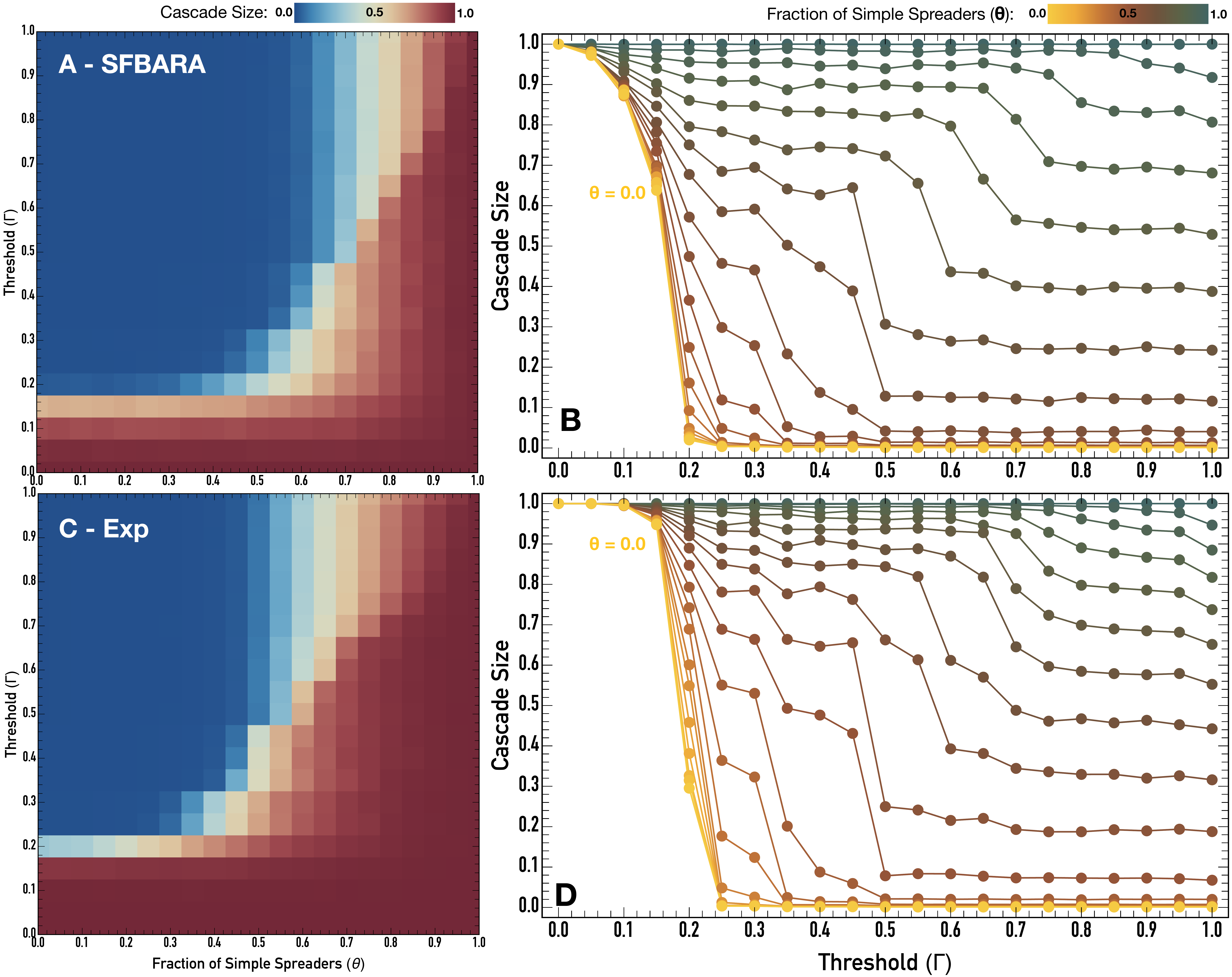}
    \caption{Threshold-based Spreaders in high degree nodes. Cascade Size as a function of Threshold-based Spreader threshold ($\Gamma$) and the fraction of Simple Spreaders ($\theta$). Panels A and C show the full picture with colors indicating no cascade (blue) or full cascade (red) parameter regions. Panels B and D show cross-sections for different values of the fraction of Simple Spreaders ($\theta$) with different colors. Other parameters are $Z = 10^3$ and $\langle k \rangle = 4$. For each condition, the reported results represent the average over $1000$ simulations for each of the $100$ network instances, totaling  $10^5$ independent simulations.}
    \label{fig2}
\end{figure*}

\subsection*{Simulations}
We start by considering a population composed by a fraction $\theta$ of Simple Spreaders and $1-\theta$ of Threshold-based Spreaders placed at random on the nodes of a network, and we simulate a cascade dynamics starting from a single randomly placed seed. We consider different thresholds ($\Gamma$) for Threshold-based Spreaders -- $0.0$ (purple), $0.25$ (blue), $0.50$ (green), $0.75$ (yellow), and $1.00$ (red) -- and study the impact on different network structures: Random (A), Exponential (B), Scale-Free (C). 

The two trivial limiting scenarios that easily verify the conditions for percolation are $\Gamma = 0$ (when Threshold-based Spreaders require only one active neighbor to become active) and $\Gamma = 1$ (when Threshold-based Spreaders never become active). Therefore, when $\Gamma = 0$, results are independent of $\theta$ (fraction of Simple Spreaders) and lead to the activation of the entire network network. In contrast, when $\Gamma = 1$, the identity line (dashed line) constitutes an upper bound of the cascade size. The cascades can be only as large as the proportion of Simple Spreaders ($\theta$) in the population since Threshold-based Spreaders never update their state. However, Threshold-Based Spreaders may disconnect the SS and not allow for a complete cascade for a low enough average degree. 

\begin{figure*}[!t]
    \centering
    \includegraphics[width=\textwidth]{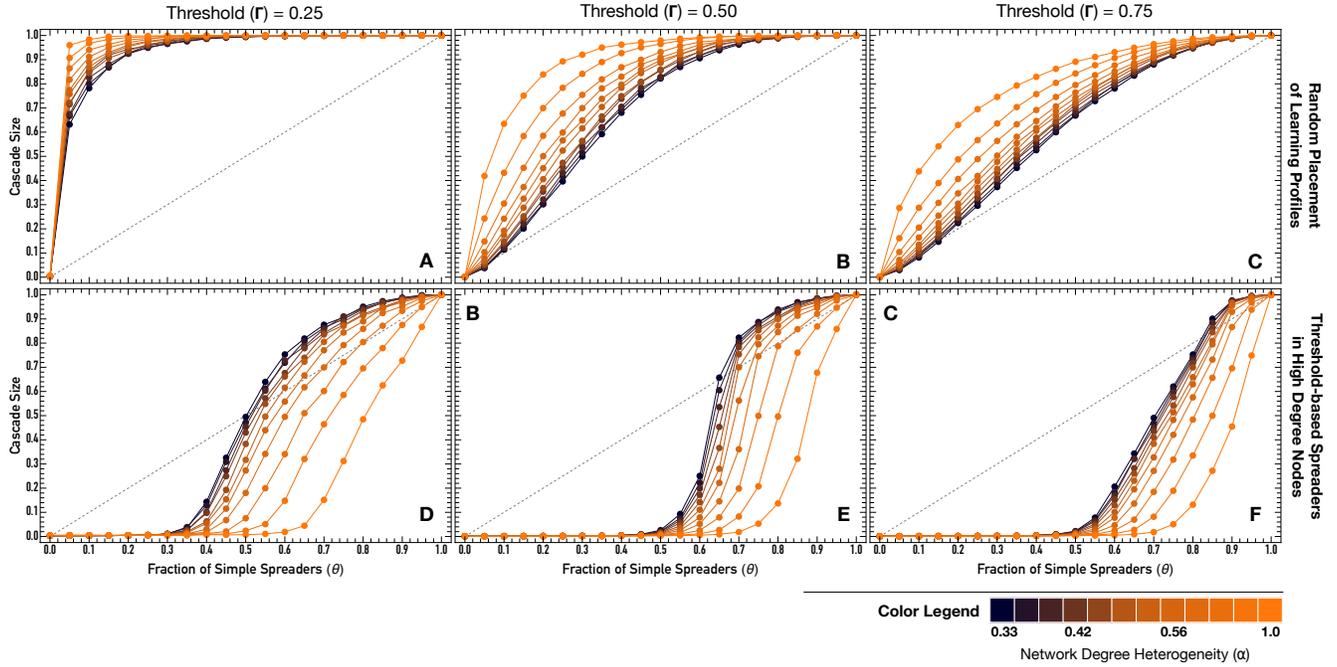}
    \caption{Effect of degree heterogeneity ($\alpha$) on Cascade Sizes as a function of the fraction of Simple Spreaders ($\theta$). The top panels (A to C) show results when Learning Profiles are placed at random and the bottom panels (D to F) when Threshold-based Spreaders are placed preferentially in high-degree nodes. The left panels (A and D) show results for $\Gamma = 0.25$, the middle panels (B and E) for $ \Gamma = 0.50$, and the right panels (C and F) for $\Gamma = 0.75$. Colors indicate the level of degree heterogeneity of the networks, which vary through the control parameter $\alpha$ used to generate the networks (See Methods). In that regard, Dark Blue colored lines represent lower degree heterogeneous networks (lower $\alpha$) and orange higher degree heterogeneous (greater $\alpha$). In each plot, the diagonal dashed line represents the identity in which an increase in the fraction of Simple Spreaders ($\theta$) would lead to an equal increase in the final fraction of activated individuals (cascade size). Other parameters are $Z = 10^3$ and $\langle k \rangle = 4$. For each condition, the reported results represent the average over $1000$ simulations for each of the $100$ network instances, totaling $10^5$ independent simulations.}
    \label{fig3}
\end{figure*}

Overall, we observe different types of curves of cascade size measured in terms of the fraction of active nodes at the end of the cascading dynamics as a function of the fraction of Simple Spreaders, depending on the threshold of Threshold-based Spreaders, network structure (degree as seen in the analytical analysis and heterogeneity), and placement of individuals. 
For most, in the scenario of very high thresholds and high fraction of Simple Spreaders (low fraction of Threshold-based Spreaders), we generally observe a linear response regarding the cascade size as a function of the fraction of Simple Spreaders, indicating that the cascade reaches only the Simple Spreaders ($\theta$). For lower thresholds, the cascade size shows some interaction between both types of individuals---cascade size is above the diagonal. 

The top panel in Figure~\ref{fig1} shows these curves when individuals are randomly placed in the network.
For heterogeneous networks, we observe mostly concave relationships, where cascades may occur even at zero fraction of Simple Spreaders and exhibit decreasing returns on the size of the cascade as we increase the fraction of Simple Spreaders. 
In the case of Random Networks, there is a critical minimum fraction of Simple Spreaders for intermediate thresholds, which increases with increasing $\Gamma$, necessary to enable a cascade. For instance, when $\Gamma = 0.25$ cascades only unfold for $\theta > 0.15$). After this fast growth, we recover the same growing concave response of the cascade size for increasing values of $\Gamma$. We refer to this as S-shaped response of the cascade curve. On closer inspection, we can see that the exponential networks already exhibit this behavior, representing a good intermediate case between random and scale-free.

\begin{figure}[!t]
    \centering
    \includegraphics[width=\columnwidth]{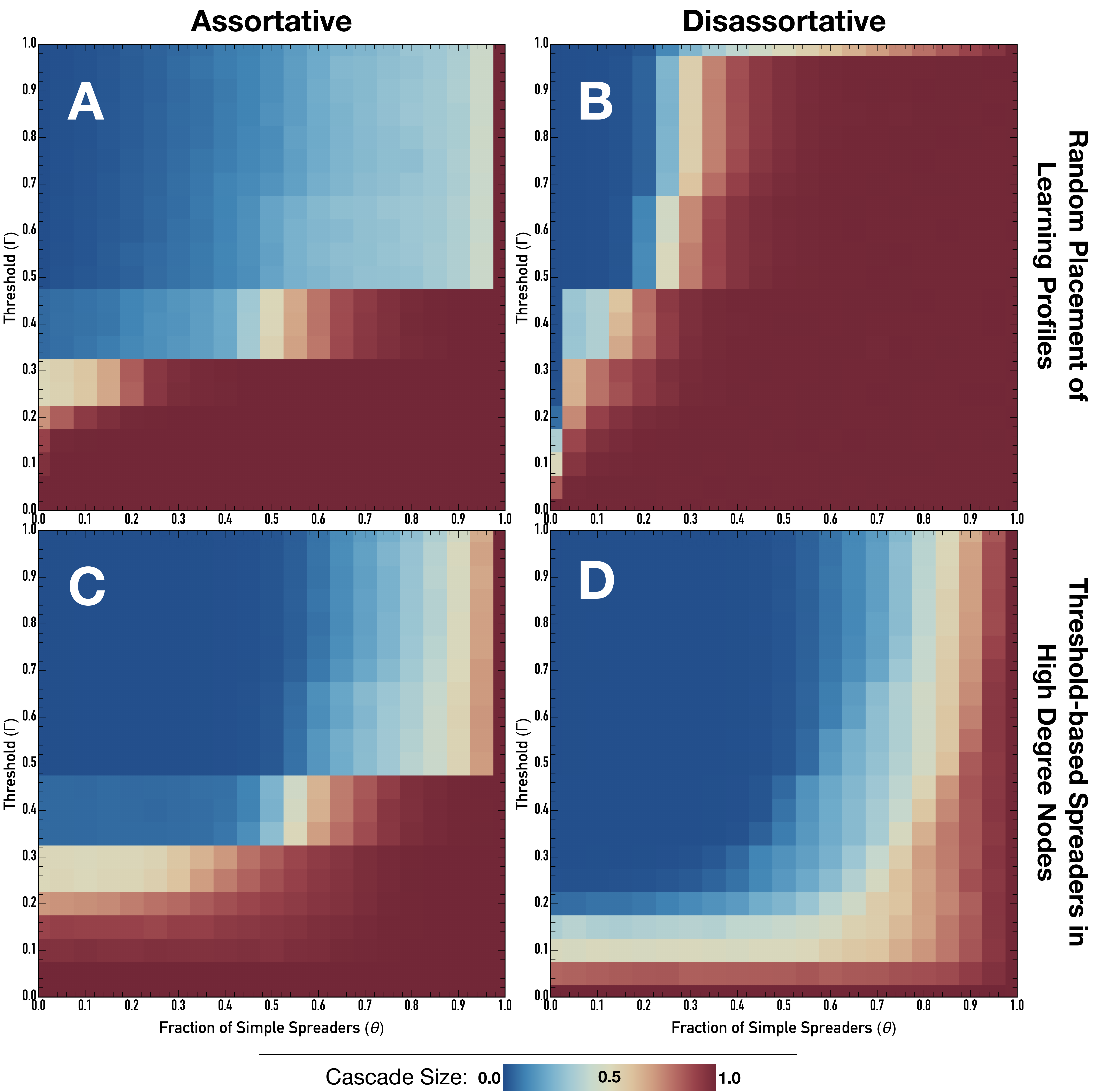}
    \caption{Cascade Size as a function of Threshold-based Spreader threshold ($\Gamma$) and the fraction of Simple Spreaders ($\theta$) on Scale Free networks with Assortative (A and C) and Disassortative (B and D) properties and when Threshold-based Spreaders are placed at random (A and B) or assorted positively with degree (C and D). Colors indicating no cascade (blue) or full cascade (red) parameter regions. Other parameters are $Z = 10^3$ and $\langle k \rangle = 4$. For each condition, the reported results represent the average over $1000$ simulations for each of the $100$ network instances, totaling  $10^5$ independent simulations.}
    \label{fig4}
\end{figure}

Even in our baseline scenario of random placement, we can already see a rich dynamical pattern due to the interplay of the different individual learning preferences and the underlying network topologies.

Next, we look at the impact of having Threshold-based spreaders preferentially assorted in nodes with higher degrees (middle panels of Figure~\ref{fig1}). We place a fraction $1-\theta$ of Threshold-based Spreaders along nodes proportionally to their degree ($k_i$). Figure~\ref{fig1} D to F results compare to those of Figure~\ref{fig1} A to C, allowing for direct assessment of the impact of placing Threshold-based Spreaders preferentially on higher degree nodes. In this case, the cascading dynamics is characterized by the S-shape response to the fraction of Simple Spreaders ($\theta$) that we saw on random networks with random placement, meaning that the cascade requires a minimum fraction of Simple Spreaders to unfold successfully. The minimal fraction increases with increasing network degree heterogeneity and $\Gamma$. For instance, in strongly degree heterogeneous networks, even a minority of 30\% Threshold-based Spreaders can impair cascade dynamics for reasonable thresholds of $\Gamma = 50\%$.

Instead, when Simple spreaders are preferentially assorted in nodes with a higher degree, Figure~\ref{fig1} G to I, we recover the baseline behavior, albeit with some small differences. For instance, Exponential networks (Figure~\ref{fig1} H) still require a critical minimum of Simple spreaders to cascade, especially for higher thresholds, similar to the results obtained for random graphs.

Overall, the suppression of the initially small cascade size with $\theta$ is observed when assorting Threshold-based spreaders preferentially on higher-degree nodes.

The cascade sizes depend additionally on the threshold that Threshold-based Spreaders use. Figure~\ref{fig2} compiles the data in Figure~\ref{fig1} to more directly show how the cascade size (in red full cascade and blue no cascade) depends on the fraction of Simple Spreaders in the population ($\theta$) and the threshold of Threshold-based Spreaders ($0 \leq \Gamma \leq 1$). The top panels show the SFBA results and the bottom panels show Exponential networks. It becomes clear that the conditions under which it is possible to observe large cascades are limited and become narrower for increasing thresholds ($\Gamma$), an outcome amplified in SFBA (high heterogeneity) compared to EXP (lower heterogeneity). Having Threshold-based Spreaders placed preferentially on high-degree nodes offers resistance to spreading and adopting information. Naturally, this can be good (if we are talking about misinformation) or bad (if we are talking about innovation).

\begin{figure}[!t]
    \centering
    \includegraphics[width=\columnwidth]{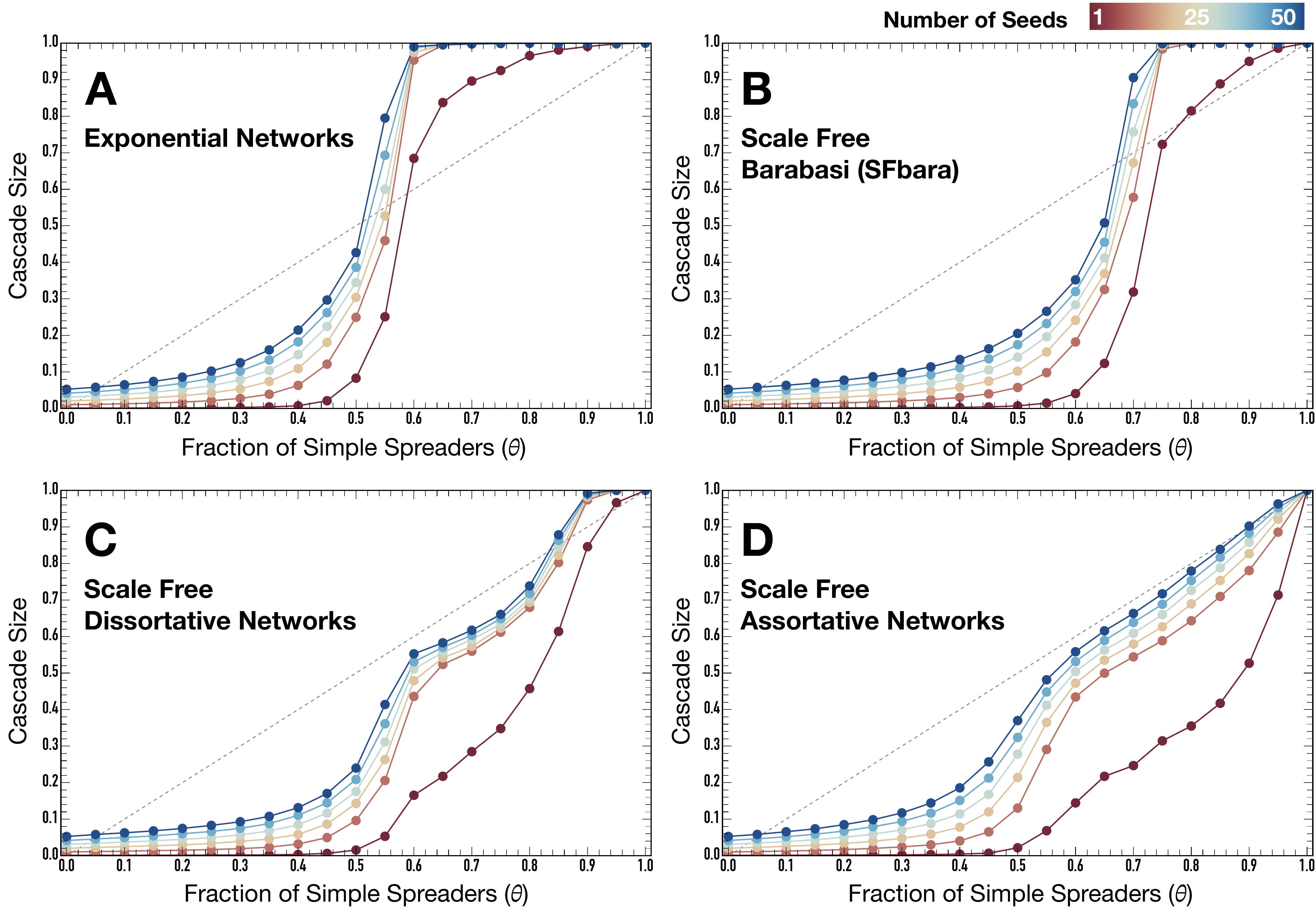}
    \caption{Cascade Sizes as a function of the fraction of Simple Spreaders and different number of seeds (colors, from dark red with one seed to dark blue with 50). Each panel represents the results for different network structures: exponential graphs (A); scale-free Barabasi-Albert (B); scale-free disassortative (C); and scale-free assortative (D). The diagonal dashed line represents the identity in which an increase in the fraction of Simple Spreaders ($\theta$) would lead to an equal increase in the final fraction of activated individuals (cascade size). Other parameters are $\Gamma = 0.5$, $Z = 10^3$ and $\langle k \rangle = 4$. For each condition, the reported results represent the average over $1000$ simulations for each of the $100$ network instances, totaling  $10^5$ independent simulations.}
    \label{fig6}
\end{figure}


From Figure~\ref{fig1}, it is clear that degree heterogeneity plays a role in the necessary conditions to generate an information cascade. Figure~\ref{fig3} explores in more detail the role of degree heterogeneity by comparing the cascade size as a function of the fraction of Simple Spreaders ($\theta$) for a set of networks that interpolate low ($\alpha = 1/3$, dark blue) and high ($\alpha = 1.0$, orange) degree heterogeneity levels, which correspond to power-law degree distributed networks with different exponents (see Methods). 

Figure~\ref{fig3} top panels show the results for Random Placement of Learning profiles and the bottom panels for the case when Threshold-based Spreaders occupy high-degree nodes preferentially. We consider also three different values of  $\Gamma = 0.25$ (A and D), $0.50$ (B and E), and $0.75$ (C and F). Results clarify two findings. First, in the bottom panels, the critical number of SS necessary to observe a cascade when Threshold-based spreaders occupy preferentially high-degree nodes increases with increasing degree heterogeneity. This is clearly visible with the right shift that is observed in the bottom curves for all values of the $\Gamma$. Secondly, the role of degree heterogeneity switches from amplifying cascades when Learning profiles are randomly placed (top panels) to suppressing of cascades when Threshold-based Spreaders are placed in high-degree nodes (bottom panels). This can be seen in Figure~\ref{fig3} by the change in the ordering of the colors in the curves. 

Past works have highlighted the role of Assortativity mixing in cascading dynamics \cite{mutlu2021effects,belhaj2016value,izquierdo2018mixing}, and we show it also plays a role in amplifying the previously discussed outcomes. Figure~\ref{fig4} shows the results obtained for assortative (panels A and C) and disassortative (panels B and D) SF networks when Threshold-based Spreaders are placed at random in the population (panels A and B) or preferentially located in higher degree nodes (panels C and D). We show that the overall dynamics of Assortative networks are relatively insensitive to the placement of Threshold-based Spreaders and usually only sensitive to the threshold of Threshold-based Spreaders ($\Gamma$).

In contrast, disassortative networks show two opposite outcomes depending on the placement of Threshold-based Spreaders. First, when Threshold-based Spreaders are randomly placed, cascades typically are large for the entire range of parameters and require a minimal frequency of Simple Spreaders to unfold. Information can spread through low-degree nodes and build influence to reach hubs. Secondly, if Threshold-based Spreaders are placed preferentially on higher-degree nodes, cascades rarely unfold unless Simple Spreaders represent more than 80\% of the population. In such cases, hubs mediate the diffusion of information between low-degree individuals. As such, they effectively block the diffusion of information, which faces challenges in building the necessary quorum to cascade.

So far, we have considered only the very strict scenario of a cascade starting from a single randomly placed seed. We tested different network topologies and, in particular, how the distribution of Learning Profiles leads to dramatically different outcomes. We now explore how sensitive the above results are to the number of seeds, and we focus on the case of Threshold-based Spreaders placed preferentially in higher-degree nodes. Such variation in the number of seeds naturally facilitates the cascade dynamics and can be useful to overcome the "cold-start" problem and the barriers Threshold-based Spreaders pose. Indeed, in the literature, this is often the point of analysis. Figure~\ref{fig6} shows the impact of seeding up to 5\% of the population (50 individuals out of 1000). We assume that seeds are placed at random.

Results show that, while increasing the number of seeds effectively increases the cascade size, and eases the conditions (e.g., critical number of Simple Spreaders), we still recover the S-shape transition---initially no or weak response, then it grows fast with $\theta$, then saturates---in Exponential and Scale-Free networks we identified with a single seed. A similar, although smoothed, dynamic is also present in both disassortative and assortative variants of scale-free networks, with little gains observed up to a threshold. In particular, assortative networks can never generate full cascades unless every individual is a Simple Spreader. This result contrasts with Exponential and Scale-Free networks where the presence of 50\% or 70\%, respectively, Simple Spreaders is a sufficient condition to generate full cascades.

\section*{\label{sec:conclusion} Discussion}
Past research has traditionally considered homogeneous populations in terms of individual adoption/learning profiles. Here, we delve into the case of a cascading dynamics process on populations that mix and controllably place Simple Spreaders and Threshold-based Spreaders. We assess how the balance of both roles and their placement on a wide range of social networks affects the cascade of novel information starting from a small number of randomly placed seeds.

We show that when both roles---Simple Spreaders and Threshold-based Spreaders---are randomly placed in a social network, they create positive synergies and result in cascades that overcome the adoption barriers of homogeneous populations of Threshold-based Spreaders. This is particularly true for degree heterogeneous networks (scale-free structures) and is somewhat more nuanced in random social structures.

However, when Threshold-based Spreaders are preferentially placed in higher-degree nodes on degree heterogeneous networks, they effectively impair the spread of information. This leads to the existence of a critical minimum number of Simple Spreaders necessary for the information to cascade. Surprisingly, these critical values are relatively large, implying that populations resist the diffusion of information that starts within the population. A dramatic outcome happens in disassortative mixing, where the mediating role of hubs completely blocks the cascade of information even when Threshold-based Spreaders have a relatively low threshold.

The implications of these results are dual and depend on whether the information being spread is socially good (e.g., innovations) or bad (e.g., misinformation). In the former case, Threshold-based Spreaders can impair the diffusion of innovations and require external intervention to facilitate transformative processes that percolate to the entire population. However, in the latter case, Simple Spreaders can represent agents placed in the social network who operate to amplify the diffusion of information and the likelihood of its virality, in particular, in contexts where a priori expectation is that information diffuses on a complex contagion scenario. In that sense, the costs do not seem to justify the means, as in many situations, well-located Threshold-based Spreaders can effectively block the cascade dynamics even in the presence of many Simple Spreaders. More importantly, it highlights the importance of having Threshold-based Spreaders influencers that operate as information filters. For practical applications, targeting high-degree nodes acting as Simple Spreaders in heterogeneous networks could suppress harmful cascades. Strategically replacing them with Threshold-based Spreaders can enhance the tunability of the spread of beneficial information. Tailoring strategies to specific network topologies, such as assortative or disassortative networks, provides a flexible toolkit for controlling cascade dynamics. For example, reducing hub connectivity or strategically placing Threshold-based Spreaders at influential positions can act as an effective barrier against misinformation. Reducing their threshold can lower that barrier. Conversely, the same interventions can inhibit the adoption of innovations, emphasizing the dual nature of these strategies.

Further exploring the role of multiple learning profiles in social systems is pertinent. Future research should revisit past paradigms in the literature on diffusion of innovations and influence maximization to benchmark how traditional seeding strategies can cope with heterogeneous populations of individuals, particularly when their distribution is structurally biased.

While our study offers valuable analytical and numerical insights, certain limitations reduce the immediate applicability of its findings to real-world networks. For instance, the empirical existence, abundance, and distribution of the different behavioral types are needed for application in specific practical settings. Similarly, the reliance on fixed-degree distributions and simplified network assumptions may overlook the complexities of real social systems, such as link dynamics, multilayered interactions, and varying connection strengths. Addressing these gaps through empirical validation, dynamic network modeling, and multilayered network representations could significantly strengthen the robustness and applicability of the results, opening pathways for future research.

\bibliography{references}

\begin{thebibliography}{10}
\urlstyle{rm}
\expandafter\ifx\csname url\endcsname\relax
  \def\url#1{\texttt{#1}}\fi
\expandafter\ifx\csname urlprefix\endcsname\relax\def\urlprefix{URL }\fi
\expandafter\ifx\csname doiprefix\endcsname\relax\def\doiprefix{DOI: }\fi
\providecommand{\bibinfo}[2]{#2}
\providecommand{\eprint}[2][]{\url{#2}}

\bibitem{zhong2017linear}
\bibinfo{author}{Zhong, Y.~D.}, \bibinfo{author}{Srivastava, V.} \&
  \bibinfo{author}{Leonard, N.~E.}
\newblock \bibinfo{title}{On the linear threshold model for diffusion of
  innovations in multiplex social networks}.
\newblock In \emph{\bibinfo{booktitle}{2017 IEEE 56th Annual Conference on
  Decision and Control (CDC)}}, \bibinfo{pages}{2593--2598}
  (\bibinfo{organization}{IEEE}, \bibinfo{year}{2017}).

\bibitem{montanari2010spread}
\bibinfo{author}{Montanari, A.} \& \bibinfo{author}{Saberi, A.}
\newblock \bibinfo{journal}{\bibinfo{title}{The spread of innovations in social
  networks}}.
\newblock {\emph{\JournalTitle{Proceedings of the National Academy of
  Sciences}}} \textbf{\bibinfo{volume}{107}}, \bibinfo{pages}{20196--20201}
  (\bibinfo{year}{2010}).

\bibitem{watts2007influentials}
\bibinfo{author}{Watts, D.~J.} \& \bibinfo{author}{Dodds, P.~S.}
\newblock \bibinfo{journal}{\bibinfo{title}{Influentials, networks, and public
  opinion formation}}.
\newblock {\emph{\JournalTitle{Journal of consumer research}}}
  \textbf{\bibinfo{volume}{34}}, \bibinfo{pages}{441--458}
  (\bibinfo{year}{2007}).

\bibitem{aral2012identifying}
\bibinfo{author}{Aral, S.} \& \bibinfo{author}{Walker, D.}
\newblock \bibinfo{journal}{\bibinfo{title}{Identifying influential and
  susceptible members of social networks}}.
\newblock {\emph{\JournalTitle{Science}}} \textbf{\bibinfo{volume}{337}},
  \bibinfo{pages}{337--341} (\bibinfo{year}{2012}).

\bibitem{molaei2018information}
\bibinfo{author}{Molaei, S.}, \bibinfo{author}{Babaei, S.},
  \bibinfo{author}{Salehi, M.} \& \bibinfo{author}{Jalili, M.}
\newblock \bibinfo{journal}{\bibinfo{title}{Information spread and topic
  diffusion in heterogeneous information networks}}.
\newblock {\emph{\JournalTitle{Scientific reports}}}
  \textbf{\bibinfo{volume}{8}}, \bibinfo{pages}{9549} (\bibinfo{year}{2018}).

\bibitem{cinelli2020covid}
\bibinfo{author}{Cinelli, M.} \emph{et~al.}
\newblock \bibinfo{journal}{\bibinfo{title}{The covid-19 social media
  infodemic}}.
\newblock {\emph{\JournalTitle{Scientific reports}}}
  \textbf{\bibinfo{volume}{10}}, \bibinfo{pages}{1--10} (\bibinfo{year}{2020}).

\bibitem{prieto2021vaccination}
\bibinfo{author}{Prieto~Curiel, R.} \&
  \bibinfo{author}{Gonz{\'a}lez~Ram{\'\i}rez, H.}
\newblock \bibinfo{journal}{\bibinfo{title}{Vaccination strategies against
  covid-19 and the diffusion of anti-vaccination views}}.
\newblock {\emph{\JournalTitle{Scientific Reports}}}
  \textbf{\bibinfo{volume}{11}}, \bibinfo{pages}{6626} (\bibinfo{year}{2021}).

\bibitem{kermani2017novel}
\bibinfo{author}{Kermani, M. A. M.~A.}, \bibinfo{author}{Ardestani, S. F.~F.},
  \bibinfo{author}{Aliahmadi, A.} \& \bibinfo{author}{Barzinpour, F.}
\newblock \bibinfo{journal}{\bibinfo{title}{A novel game theoretic approach for
  modeling competitive information diffusion in social networks with
  heterogeneous nodes}}.
\newblock {\emph{\JournalTitle{Physica A: statistical mechanics and its
  applications}}} \textbf{\bibinfo{volume}{466}}, \bibinfo{pages}{570--582}
  (\bibinfo{year}{2017}).

\bibitem{shao2017spread}
\bibinfo{author}{Shao, C.}, \bibinfo{author}{Ciampaglia, G.~L.},
  \bibinfo{author}{Varol, O.}, \bibinfo{author}{Flammini, A.} \&
  \bibinfo{author}{Menczer, F.}
\newblock \bibinfo{journal}{\bibinfo{title}{The spread of misinformation by
  social bots}}.
\newblock {\emph{\JournalTitle{arXiv preprint arXiv:1707.07592}}}
  (\bibinfo{year}{2017}).

\bibitem{shao2018spread}
\bibinfo{author}{Shao, C.} \emph{et~al.}
\newblock \bibinfo{journal}{\bibinfo{title}{The spread of low-credibility
  content by social bots}}.
\newblock {\emph{\JournalTitle{Nature communications}}}
  \textbf{\bibinfo{volume}{9}}, \bibinfo{pages}{1--9} (\bibinfo{year}{2018}).

\bibitem{rossi2020detecting}
\bibinfo{author}{Rossi, S.}, \bibinfo{author}{Rossi, M.},
  \bibinfo{author}{Upreti, B.~R.} \& \bibinfo{author}{Liu, Y.}
\newblock \bibinfo{title}{Detecting political bots on twitter during the 2019
  finnish parliamentary election}.
\newblock In \emph{\bibinfo{booktitle}{Annual Hawaii International Conference
  on System Sciences}}, \bibinfo{pages}{2430--2439}
  (\bibinfo{organization}{Hawaii International Conference on System Sciences},
  \bibinfo{year}{2020}).

\bibitem{bradshaw2017troops}
\bibinfo{author}{Bradshaw, S.} \& \bibinfo{author}{Howard, P.}
\newblock \bibinfo{journal}{\bibinfo{title}{Troops, trolls and troublemakers: A
  global inventory of organized social media manipulation}}.
\newblock {\emph{\JournalTitle{Computational Propaganda Research Project}}}
  (\bibinfo{year}{2017}).

\bibitem{shahid2022you}
\bibinfo{author}{Shahid, W.} \emph{et~al.}
\newblock \bibinfo{journal}{\bibinfo{title}{Are you a cyborg, bot or human?—a
  survey on detecting fake news spreaders}}.
\newblock {\emph{\JournalTitle{IEEE Access}}} \textbf{\bibinfo{volume}{10}},
  \bibinfo{pages}{27069--27083} (\bibinfo{year}{2022}).

\bibitem{stein2023network}
\bibinfo{author}{Stein, J.}, \bibinfo{author}{Keuschnigg, M.} \&
  \bibinfo{author}{van~de Rijt, A.}
\newblock \bibinfo{journal}{\bibinfo{title}{Network segregation and the
  propagation of misinformation}}.
\newblock {\emph{\JournalTitle{Scientific Reports}}}
  \textbf{\bibinfo{volume}{13}}, \bibinfo{pages}{917} (\bibinfo{year}{2023}).

\bibitem{zheng2012social}
\bibinfo{author}{Zheng, X.}, \bibinfo{author}{Zhong, Y.},
  \bibinfo{author}{Zeng, D.} \& \bibinfo{author}{Wang, F.-Y.}
\newblock \bibinfo{journal}{\bibinfo{title}{Social influence and spread
  dynamics in social networks}}.
\newblock {\emph{\JournalTitle{Frontiers of Computer Science}}}
  \textbf{\bibinfo{volume}{6}}, \bibinfo{pages}{611--620}
  (\bibinfo{year}{2012}).

\bibitem{cencetti2023distinguishing}
\bibinfo{author}{Cencetti, G.}, \bibinfo{author}{Contreras, D.~A.},
  \bibinfo{author}{Mancastroppa, M.} \& \bibinfo{author}{Barrat, A.}
\newblock \bibinfo{journal}{\bibinfo{title}{Distinguishing simple and complex
  contagion processes on networks}}.
\newblock {\emph{\JournalTitle{Physical Review Letters}}}
  \textbf{\bibinfo{volume}{130}}, \bibinfo{pages}{247401}
  (\bibinfo{year}{2023}).

\bibitem{zanette2002dynamics}
\bibinfo{author}{Zanette, D.~H.}
\newblock \bibinfo{journal}{\bibinfo{title}{Dynamics of rumor propagation on
  small-world networks}}.
\newblock {\emph{\JournalTitle{Physical review E}}}
  \textbf{\bibinfo{volume}{65}}, \bibinfo{pages}{041908}
  (\bibinfo{year}{2002}).

\bibitem{dodds2004universal}
\bibinfo{author}{Dodds, P.~S.} \& \bibinfo{author}{Watts, D.~J.}
\newblock \bibinfo{journal}{\bibinfo{title}{Universal behavior in a generalized
  model of contagion}}.
\newblock {\emph{\JournalTitle{Physical review letters}}}
  \textbf{\bibinfo{volume}{92}}, \bibinfo{pages}{218701}
  (\bibinfo{year}{2004}).

\bibitem{smolyak2020mitigation}
\bibinfo{author}{Smolyak, A.}, \bibinfo{author}{Levy, O.},
  \bibinfo{author}{Vodenska, I.}, \bibinfo{author}{Buldyrev, S.} \&
  \bibinfo{author}{Havlin, S.}
\newblock \bibinfo{journal}{\bibinfo{title}{Mitigation of cascading failures in
  complex networks}}.
\newblock {\emph{\JournalTitle{Scientific reports}}}
  \textbf{\bibinfo{volume}{10}}, \bibinfo{pages}{16124} (\bibinfo{year}{2020}).

\bibitem{vespignani2012modelling}
\bibinfo{author}{Vespignani, A.}
\newblock \bibinfo{journal}{\bibinfo{title}{Modelling dynamical processes in
  complex socio-technical systems}}.
\newblock {\emph{\JournalTitle{Nature physics}}} \textbf{\bibinfo{volume}{8}},
  \bibinfo{pages}{32--39} (\bibinfo{year}{2012}).

\bibitem{sood2008voter}
\bibinfo{author}{Sood, V.}, \bibinfo{author}{Antal, T.} \&
  \bibinfo{author}{Redner, S.}
\newblock \bibinfo{journal}{\bibinfo{title}{Voter models on heterogeneous
  networks}}.
\newblock {\emph{\JournalTitle{Physical Review E}}}
  \textbf{\bibinfo{volume}{77}}, \bibinfo{pages}{041121}
  (\bibinfo{year}{2008}).

\bibitem{sood2005voter}
\bibinfo{author}{Sood, V.} \& \bibinfo{author}{Redner, S.}
\newblock \bibinfo{journal}{\bibinfo{title}{Voter model on heterogeneous
  graphs}}.
\newblock {\emph{\JournalTitle{Physical review letters}}}
  \textbf{\bibinfo{volume}{94}}, \bibinfo{pages}{178701}
  (\bibinfo{year}{2005}).

\bibitem{kempe2003maximizing}
\bibinfo{author}{Kempe, D.}, \bibinfo{author}{Kleinberg, J.} \&
  \bibinfo{author}{Tardos, {\'E}.}
\newblock \bibinfo{title}{Maximizing the spread of influence through a social
  network}.
\newblock In \emph{\bibinfo{booktitle}{Proceedings of the ninth ACM SIGKDD
  international conference on Knowledge discovery and data mining}},
  \bibinfo{pages}{137--146} (\bibinfo{year}{2003}).

\bibitem{karsai2016local}
\bibinfo{author}{Karsai, M.}, \bibinfo{author}{I{\~n}iguez, G.},
  \bibinfo{author}{Kikas, R.}, \bibinfo{author}{Kaski, K.} \&
  \bibinfo{author}{Kert{\'e}sz, J.}
\newblock \bibinfo{journal}{\bibinfo{title}{Local cascades induced global
  contagion: How heterogeneous thresholds, exogenous effects, and unconcerned
  behaviour govern online adoption spreading}}.
\newblock {\emph{\JournalTitle{Scientific reports}}}
  \textbf{\bibinfo{volume}{6}}, \bibinfo{pages}{27178} (\bibinfo{year}{2016}).

\bibitem{centola2007cascade}
\bibinfo{author}{Centola, D.}, \bibinfo{author}{Egu{\'\i}luz, V.~M.} \&
  \bibinfo{author}{Macy, M.~W.}
\newblock \bibinfo{journal}{\bibinfo{title}{Cascade dynamics of complex
  propagation}}.
\newblock {\emph{\JournalTitle{Physica A: Statistical Mechanics and its
  Applications}}} \textbf{\bibinfo{volume}{374}}, \bibinfo{pages}{449--456}
  (\bibinfo{year}{2007}).

\bibitem{centola2010spread}
\bibinfo{author}{Centola, D.}
\newblock \bibinfo{journal}{\bibinfo{title}{The spread of behavior in an online
  social network experiment}}.
\newblock {\emph{\JournalTitle{science}}} \textbf{\bibinfo{volume}{329}},
  \bibinfo{pages}{1194--1197} (\bibinfo{year}{2010}).

\bibitem{centola2018behavior}
\bibinfo{author}{Centola, D.}
\newblock \emph{\bibinfo{title}{How behavior spreads: The science of complex
  contagions}}, vol.~\bibinfo{volume}{3} (\bibinfo{publisher}{Princeton
  University Press Princeton, NJ}, \bibinfo{year}{2018}).

\bibitem{derechin2023cascades}
\bibinfo{author}{Derechin, J.}
\newblock \bibinfo{journal}{\bibinfo{title}{Cascades in capacity constrained
  agents}}.
\newblock {\emph{\JournalTitle{Plos one}}} \textbf{\bibinfo{volume}{18}},
  \bibinfo{pages}{e0280326} (\bibinfo{year}{2023}).

\bibitem{borges2024social}
\bibinfo{author}{Borges, H.~M.}, \bibinfo{author}{Vasconcelos, V.~V.} \&
  \bibinfo{author}{Pinheiro, F.~L.}
\newblock \bibinfo{journal}{\bibinfo{title}{How social rewiring preferences
  bridge polarized communities}}.
\newblock {\emph{\JournalTitle{Chaos, Solitons \& Fractals}}}
  \textbf{\bibinfo{volume}{180}}, \bibinfo{pages}{114594}
  (\bibinfo{year}{2024}).

\bibitem{vasconcelos2019consensus}
\bibinfo{author}{Vasconcelos, V.~V.}, \bibinfo{author}{Levin, S.~A.} \&
  \bibinfo{author}{Pinheiro, F.~L.}
\newblock \bibinfo{journal}{\bibinfo{title}{Consensus and polarization in
  competing complex contagion processes}}.
\newblock {\emph{\JournalTitle{Journal of the Royal Society Interface}}}
  \textbf{\bibinfo{volume}{16}}, \bibinfo{pages}{20190196}
  (\bibinfo{year}{2019}).

\bibitem{berger2008identity}
\bibinfo{author}{Berger, J.}
\newblock \bibinfo{journal}{\bibinfo{title}{Identity signaling, social
  influence, and social contagion}}.
\newblock {\emph{\JournalTitle{Understanding peer influence in children and
  adolescents}}} \bibinfo{pages}{181--199} (\bibinfo{year}{2008}).

\bibitem{karsai2014complex}
\bibinfo{author}{Karsai, M.}, \bibinfo{author}{Iniguez, G.},
  \bibinfo{author}{Kaski, K.} \& \bibinfo{author}{Kert{\'e}sz, J.}
\newblock \bibinfo{journal}{\bibinfo{title}{Complex contagion process in
  spreading of online innovation}}.
\newblock {\emph{\JournalTitle{Journal of The Royal Society Interface}}}
  \textbf{\bibinfo{volume}{11}}, \bibinfo{pages}{20140694}
  (\bibinfo{year}{2014}).

\bibitem{young2009innovation}
\bibinfo{author}{Young, H.~P.}
\newblock \bibinfo{journal}{\bibinfo{title}{Innovation diffusion in
  heterogeneous populations: Contagion, social influence, and social
  learning}}.
\newblock {\emph{\JournalTitle{American economic review}}}
  \textbf{\bibinfo{volume}{99}}, \bibinfo{pages}{1899--1924}
  (\bibinfo{year}{2009}).

\bibitem{sprague2017evidence}
\bibinfo{author}{Sprague, D.~A.} \& \bibinfo{author}{House, T.}
\newblock \bibinfo{journal}{\bibinfo{title}{Evidence for complex contagion
  models of social contagion from observational data}}.
\newblock {\emph{\JournalTitle{PloS one}}} \textbf{\bibinfo{volume}{12}},
  \bibinfo{pages}{e0180802} (\bibinfo{year}{2017}).

\bibitem{tump2020wise}
\bibinfo{author}{Tump, A.~N.}, \bibinfo{author}{Pleskac, T.~J.} \&
  \bibinfo{author}{Kurvers, R.~H.}
\newblock \bibinfo{journal}{\bibinfo{title}{Wise or mad crowds? the cognitive
  mechanisms underlying information cascades}}.
\newblock {\emph{\JournalTitle{Science Advances}}}
  \textbf{\bibinfo{volume}{6}}, \bibinfo{pages}{eabb0266}
  (\bibinfo{year}{2020}).

\bibitem{mittal2024anti}
\bibinfo{author}{Mittal, D.}, \bibinfo{author}{Constantino, S.~M.} \&
  \bibinfo{author}{Vasconcelos, V.~V.}
\newblock \bibinfo{journal}{\bibinfo{title}{Anticonformists catalyze societal
  transitions and facilitate the expression of evolving preferences}}.
\newblock {\emph{\JournalTitle{PNAS Nexus}}} \bibinfo{pages}{pgae302}
  (\bibinfo{year}{2024}).

\bibitem{robertson1967process}
\bibinfo{author}{Robertson, T.~S.}
\newblock \bibinfo{journal}{\bibinfo{title}{The process of innovation and the
  diffusion of innovation}}.
\newblock {\emph{\JournalTitle{Journal of marketing}}}
  \textbf{\bibinfo{volume}{31}}, \bibinfo{pages}{14--19}
  (\bibinfo{year}{1967}).

\bibitem{rogers2010diffusion}
\bibinfo{author}{Rogers, E.~M.}
\newblock \emph{\bibinfo{title}{Diffusion of innovations}}
  (\bibinfo{publisher}{Simon and Schuster}, \bibinfo{year}{2010}).

\bibitem{rogers2014diffusion}
\bibinfo{author}{Rogers, E.~M.}, \bibinfo{author}{Singhal, A.} \&
  \bibinfo{author}{Quinlan, M.~M.}
\newblock \bibinfo{title}{Diffusion of innovations}.
\newblock In \emph{\bibinfo{booktitle}{An integrated approach to communication
  theory and research}}, \bibinfo{pages}{432--448}
  (\bibinfo{publisher}{Routledge}, \bibinfo{year}{2014}).

\bibitem{kleinberg2007cascading}
\bibinfo{author}{Kleinberg, J.} \emph{et~al.}
\newblock \bibinfo{journal}{\bibinfo{title}{Cascading behavior in networks:
  Algorithmic and economic issues}}.
\newblock {\emph{\JournalTitle{Algorithmic game theory}}}
  \textbf{\bibinfo{volume}{24}}, \bibinfo{pages}{613--632}
  (\bibinfo{year}{2007}).

\bibitem{karampourniotis2015impact}
\bibinfo{author}{Karampourniotis, P.~D.}, \bibinfo{author}{Sreenivasan, S.},
  \bibinfo{author}{Szymanski, B.~K.} \& \bibinfo{author}{Korniss, G.}
\newblock \bibinfo{journal}{\bibinfo{title}{The impact of heterogeneous
  thresholds on social contagion with multiple initiators}}.
\newblock {\emph{\JournalTitle{PloS one}}} \textbf{\bibinfo{volume}{10}},
  \bibinfo{pages}{e0143020} (\bibinfo{year}{2015}).

\bibitem{izquierdo2018mixing}
\bibinfo{author}{Izquierdo, S.~S.}, \bibinfo{author}{Izquierdo, L.~R.} \&
  \bibinfo{author}{L{\'o}pez-Pintado, D.}
\newblock \bibinfo{journal}{\bibinfo{title}{Mixing and diffusion in a two-type
  population}}.
\newblock {\emph{\JournalTitle{Royal Society Open Science}}}
  \textbf{\bibinfo{volume}{5}}, \bibinfo{pages}{172102} (\bibinfo{year}{2018}).

\bibitem{guilbeault2021topological}
\bibinfo{author}{Guilbeault, D.} \& \bibinfo{author}{Centola, D.}
\newblock \bibinfo{journal}{\bibinfo{title}{Topological measures for
  identifying and predicting the spread of complex contagions}}.
\newblock {\emph{\JournalTitle{Nature communications}}}
  \textbf{\bibinfo{volume}{12}}, \bibinfo{pages}{4430} (\bibinfo{year}{2021}).

\bibitem{kim2014ct}
\bibinfo{author}{Kim, J.}, \bibinfo{author}{Lee, W.} \& \bibinfo{author}{Yu,
  H.}
\newblock \bibinfo{journal}{\bibinfo{title}{Ct-ic: Continuously activated and
  time-restricted independent cascade model for viral marketing}}.
\newblock {\emph{\JournalTitle{Knowledge-Based Systems}}}
  \textbf{\bibinfo{volume}{62}}, \bibinfo{pages}{57--68}
  (\bibinfo{year}{2014}).

\bibitem{tu2022viral}
\bibinfo{author}{Tu, S.} \& \bibinfo{author}{Neumann, S.}
\newblock \bibinfo{title}{A viral marketing-based model for opinion dynamics in
  online social networks}.
\newblock In \emph{\bibinfo{booktitle}{Proceedings of the ACM Web Conference
  2022}}, \bibinfo{pages}{1570--1578} (\bibinfo{year}{2022}).

\bibitem{leskovec2007dynamics}
\bibinfo{author}{Leskovec, J.}, \bibinfo{author}{Adamic, L.~A.} \&
  \bibinfo{author}{Huberman, B.~A.}
\newblock \bibinfo{journal}{\bibinfo{title}{The dynamics of viral marketing}}.
\newblock {\emph{\JournalTitle{ACM Transactions on the Web (TWEB)}}}
  \textbf{\bibinfo{volume}{1}}, \bibinfo{pages}{5--es} (\bibinfo{year}{2007}).

\bibitem{aral2011commentary}
\bibinfo{author}{Aral, S.}
\newblock \bibinfo{journal}{\bibinfo{title}{Commentary—identifying social
  influence: A comment on opinion leadership and social contagion in new
  product diffusion}}.
\newblock {\emph{\JournalTitle{Marketing Science}}}
  \textbf{\bibinfo{volume}{30}}, \bibinfo{pages}{217--223}
  (\bibinfo{year}{2011}).

\bibitem{aral2018social}
\bibinfo{author}{Aral, S.} \& \bibinfo{author}{Dhillon, P.~S.}
\newblock \bibinfo{journal}{\bibinfo{title}{Social influence maximization under
  empirical influence models}}.
\newblock {\emph{\JournalTitle{Nature human behaviour}}}
  \textbf{\bibinfo{volume}{2}}, \bibinfo{pages}{375--382}
  (\bibinfo{year}{2018}).

\bibitem{min2018competing}
\bibinfo{author}{Min, B.} \& \bibinfo{author}{San~Miguel, M.}
\newblock \bibinfo{journal}{\bibinfo{title}{Competing contagion processes:
  Complex contagion triggered by simple contagion}}.
\newblock {\emph{\JournalTitle{Scientific reports}}}
  \textbf{\bibinfo{volume}{8}}, \bibinfo{pages}{10422} (\bibinfo{year}{2018}).

\bibitem{prier2020commanding}
\bibinfo{author}{Prier, J.}
\newblock \bibinfo{title}{Commanding the trend: Social media as information
  warfare}.
\newblock In \emph{\bibinfo{booktitle}{Information warfare in the age of cyber
  conflict}}, \bibinfo{pages}{88--113} (\bibinfo{publisher}{Routledge},
  \bibinfo{year}{2020}).

\bibitem{ferrara2020misinformation}
\bibinfo{author}{Ferrara, E.}, \bibinfo{author}{Cresci, S.} \&
  \bibinfo{author}{Luceri, L.}
\newblock \bibinfo{journal}{\bibinfo{title}{Misinformation, manipulation, and
  abuse on social media in the era of covid-19}}.
\newblock {\emph{\JournalTitle{Journal of Computational Social Science}}}
  \textbf{\bibinfo{volume}{3}}, \bibinfo{pages}{271--277}
  (\bibinfo{year}{2020}).

\bibitem{erdHos1960evolution}
\bibinfo{author}{Erd{\H{o}}s, P.}, \bibinfo{author}{R{\'e}nyi, A.}
  \emph{et~al.}
\newblock \bibinfo{journal}{\bibinfo{title}{On the evolution of random
  graphs}}.
\newblock {\emph{\JournalTitle{Publ. math. inst. hung. acad. sci}}}
  \textbf{\bibinfo{volume}{5}}, \bibinfo{pages}{17--60} (\bibinfo{year}{1960}).

\bibitem{albert2002statistical}
\bibinfo{author}{Albert, R.} \& \bibinfo{author}{Barab{\'a}si, A.-L.}
\newblock \bibinfo{journal}{\bibinfo{title}{Statistical mechanics of complex
  networks}}.
\newblock {\emph{\JournalTitle{Reviews of modern physics}}}
  \textbf{\bibinfo{volume}{74}}, \bibinfo{pages}{47} (\bibinfo{year}{2002}).

\bibitem{santos2012role}
\bibinfo{author}{Santos, F.~C.}, \bibinfo{author}{Pinheiro, F.~L.},
  \bibinfo{author}{Lenaerts, T.} \& \bibinfo{author}{Pacheco, J.~M.}
\newblock \bibinfo{journal}{\bibinfo{title}{The role of diversity in the
  evolution of cooperation}}.
\newblock {\emph{\JournalTitle{Journal of theoretical biology}}}
  \textbf{\bibinfo{volume}{299}}, \bibinfo{pages}{88--96}
  (\bibinfo{year}{2012}).

\bibitem{gleeson2008cascades}
\bibinfo{author}{Gleeson, J.~P.}
\newblock \bibinfo{journal}{\bibinfo{title}{Cascades on correlated and modular
  random networks}}.
\newblock {\emph{\JournalTitle{Physical Review E}}}
  \textbf{\bibinfo{volume}{77}}, \bibinfo{pages}{046117}
  (\bibinfo{year}{2008}).

\bibitem{xulvi2004reshuffling}
\bibinfo{author}{Xulvi-Brunet, R.} \& \bibinfo{author}{Sokolov, I.~M.}
\newblock \bibinfo{journal}{\bibinfo{title}{Reshuffling scale-free networks:
  From random to assortative}}.
\newblock {\emph{\JournalTitle{Physical Review E}}}
  \textbf{\bibinfo{volume}{70}}, \bibinfo{pages}{066102}
  (\bibinfo{year}{2004}).

\bibitem{fortunato2006scale}
\bibinfo{author}{Fortunato, S.}, \bibinfo{author}{Flammini, A.} \&
  \bibinfo{author}{Menczer, F.}
\newblock \bibinfo{journal}{\bibinfo{title}{Scale-free network growth by
  ranking}}.
\newblock {\emph{\JournalTitle{Physical review letters}}}
  \textbf{\bibinfo{volume}{96}}, \bibinfo{pages}{218701}
  (\bibinfo{year}{2006}).

\bibitem{goh2001universal}
\bibinfo{author}{Goh, K.-I.}, \bibinfo{author}{Kahng, B.} \&
  \bibinfo{author}{Kim, D.}
\newblock \bibinfo{journal}{\bibinfo{title}{Universal behavior of load
  distribution in scale-free networks}}.
\newblock {\emph{\JournalTitle{Physical review letters}}}
  \textbf{\bibinfo{volume}{87}}, \bibinfo{pages}{278701}
  (\bibinfo{year}{2001}).

\bibitem{pinheiro2017intermediate}
\bibinfo{author}{Pinheiro, F.~L.} \& \bibinfo{author}{Hartmann, D.}
\newblock \bibinfo{journal}{\bibinfo{title}{Intermediate levels of network
  heterogeneity provide the best evolutionary outcomes}}.
\newblock {\emph{\JournalTitle{Scientific reports}}}
  \textbf{\bibinfo{volume}{7}}, \bibinfo{pages}{15242} (\bibinfo{year}{2017}).

\bibitem{jalili2017information}
\bibinfo{author}{Jalili, M.} \& \bibinfo{author}{Perc, M.}
\newblock \bibinfo{journal}{\bibinfo{title}{Information cascades in complex
  networks}}.
\newblock {\emph{\JournalTitle{Journal of Complex Networks}}}
  \textbf{\bibinfo{volume}{5}}, \bibinfo{pages}{665--693}
  (\bibinfo{year}{2017}).

\bibitem{molloy1995critical}
\bibinfo{author}{Molloy, M.} \& \bibinfo{author}{Reed, B.}
\newblock \bibinfo{journal}{\bibinfo{title}{A critical point for random graphs
  with a given degree sequence}}.
\newblock {\emph{\JournalTitle{Random structures \& algorithms}}}
  \textbf{\bibinfo{volume}{6}}, \bibinfo{pages}{161--180}
  (\bibinfo{year}{1995}).

\bibitem{mutlu2021effects}
\bibinfo{author}{Mutlu, E.} \& \bibinfo{author}{Garibay, O.~O.}
\newblock \bibinfo{title}{Effects of assortativity on consensus formation with
  heterogeneous agents}.
\newblock In \emph{\bibinfo{booktitle}{Conference of the Computational Social
  Science Society of the Americas}}, \bibinfo{pages}{1--10}
  (\bibinfo{organization}{Springer}, \bibinfo{year}{2021}).

\bibitem{belhaj2016value}
\bibinfo{author}{Belhaj, M.} \& \bibinfo{author}{Dero{\"\i}an, F.}
\newblock \bibinfo{journal}{\bibinfo{title}{The value of network information:
  Assortative mixing makes the difference}}.
\newblock {\emph{\JournalTitle{Games and Economic Behavior}}}
  \textbf{\bibinfo{volume}{126}}, \bibinfo{pages}{428--442}
  (\bibinfo{year}{2021}).

\end{thebibliography}

\section*{Acknowledgements}
FLP acknowledges the financial support provided by FCT Portugal under the project UIDB/04152/2020 -- Centro de Investigação em Gestão de Informação (MagIC). VVV acknowledges funding from ENLENS under the project "The Cost of Large-Scale Transitions: Introducing Effective Targeted Incentives."

\section*{Author contributions statement}
FLP and VVV contributed equally to the elaboration of this manuscript.

\section*{Competing interests statement}
The authors declare no competing interests.

\section*{Data Availability Statement}
The datasets generated during and/or analyzed during the current study along with the code used to carry out the simulations are available from the corresponding author on reasonable request.

\end{document}